\documentclass[notitlepage,12pt]{jedm}
\usepackage[table]{xcolor}
\usepackage{graphicx}
\graphicspath{ {./images/} }
\usepackage{url}
\usepackage{hyperref}
\hypersetup{
  colorlinks   = true, 
  urlcolor     = blue, 
  linkcolor    = blue, 
  citecolor   = blue 
}
\usepackage{subfigure}
\usepackage{amsmath}
\usepackage[british]{babel}
\usepackage{hhline}
\usepackage{multirow}
\usepackage{dsfont}

\begin{document}

\title{ClickTree: A Tree-based Method for Predicting Math Students' Performance Based on Clickstream Data}
\date{} 

\author{{\large Narjes Rohani}\\University of Edinburgh\\Narjes.Rohani@ed.ac.uk \and {\large Behnam Rohani}\\Sharif University of Technology\\behnam.rohani058@sharif.edu
\and {\large Areti Manataki}\\University of St Andrews\\A.Manataki@st-andrews.ac.uk}
\maketitle

\begin{abstract}
The prediction of student performance and the analysis of students' learning behavior play an important role in enhancing online courses. By analysing a massive amount of clickstream data that captures student behavior, educators can gain valuable insights into the factors that influence academic outcomes and identify areas of improvement in courses. 

In this study, we developed ClickTree, a tree-based methodology, to predict student performance in mathematical assignments based on students' clickstream data. We extracted a set of features, including problem-level, assignment-level and student-level features, from the extensive clickstream data and trained a CatBoost tree to predict whether a student successfully answers a problem in an assignment. The developed method achieved an AUC of 0.78844 in the Educational Data Mining Cup 2023 and ranked second in the competition.

Furthermore, our results indicate that students encounter more difficulties in the problem types that they must select a subset of answers from a given set as well as problem subjects of Algebra II. Additionally, students who performed well in answering end-unit assignment problems engaged more with in-unit assignments and answered more problems correctly, while those who struggled had higher tutoring request rate. The proposed method can be utilized to improve students' learning experiences, and the above insights can be integrated into mathematical courses to enhance students' learning outcomes.
\newline
{\parindent0pt
\textbf{Keywords:} Student performance prediction, Educational data mining, Mathematics, Learning behaviour, Learning analytics.
}
\end{abstract}

\section{Introduction}
In recent years, massive amounts of log data have been collected from students' interactions with online courses, providing researchers with valuable information to analyze student behavior and its impact on academic performance \cite{yi2018predictive,aljohani2019predicting}. By examining clickstream data, educators can gain deeper insights into students' study habits, navigation patterns, and levels of engagement \cite{wen2014identifying,li2020using,matcha2020analytics}. This knowledge helps identify areas where students may be struggling or disengaged, allowing for timely intervention and personalized support \cite{matcha2020analytics,matcha2019analytics}. Moreover, clickstream data analysis enables educators to identify effective learning patterns and resources that influence student performance \cite{aljohani2019predicting}.

By harnessing the power of educational data mining and artificial intelligence, teachers and educators can make informed decisions for optimizing teaching methodologies, enhancing the learning experience, and ultimately improving student outcomes \cite{schumacher2018features,jang2022practical,koedinger2013new}. In line with this objective, the Educational Data Mining Cup (EDMcup) 2023 \cite{edm-cup-2023} was launched as a competition to predict students' end-unit-assignment scores in math problems collected from the ASSISTments online platform \cite{heffernan2014assistments}. The EDMcup provided access to clickstream data, including millions of student actions, as well as curriculum data, to predict students' end-unit-assignment scores based on their behavior during in-unit assignments \cite{edm-cup-2023}.

The literature review reveals two types of performance prediction tasks: program-level and course-level \cite{cui2019predictive,namoun2020predicting,liu2022predicting}. Program-level tasks focus on predicting student dropouts or graduation probabilities from a degree program, while course-level tasks involve predicting students' scores, grades, or pass/fail status in specific courses \cite{lemay2022predicting}. Course-level prediction tasks often aim to identify at-risk students who are more likely to fail and provide timely interventions to help them succeed \cite{akccapinar2019using,akram2019predicting}. Predictive analysis at the course level offers valuable insights for enhancing teaching and learning, enabling instructors to understand student behavior, design effective instruction, and provide targeted support \cite{akram2019predicting,oliva2021learning}. 

Various data mining algorithms, such as Decision Trees, Support Vector Machines, and Random Forest, are commonly employed in the educational data mining field \cite{namoun2020predicting,lopez2021early}. In recent years, the utilization of clickstream data, which captures students' online learning behaviors, such as accessing resources and completing assignments, has gained popularity in performance prediction research \cite{yurum2023use,baker2020benefits}. However, extracting meaningful features from raw clickstream data remains a challenge \cite{edm-cup-2023,baker2020benefits}, and further research is needed to process the large amount of raw data on student behavior and convert it into meaningful features that can accurately predict student outcomes.

This study introduces ClickTree, a tree-based model employing clickstream data, for predicting students' scores on each mathematical problem of end-unit assignments based on their behavior during in-unit assignments. Millions of student actions on the ASSISTments platform were analyzed, and a set of meaningful features at the student, assignment, and problem levels were extracted to predict student performance accurately. The CatBoost tree algorithm was applied to the extracted features, and the model achieved an AUC score of approximately 79\%, ranking second in the EDMcup2023, and thus demonstrating its potential for accurately predicting students' outcomes in math courses. Additionally, we explored students' outcomes for different problem types and topics to identify the most challenging one. Furthermore, learning behaviors of successful students were compared to those of struggling students to uncover behavioral patterns between the two groups.


The research questions investigated in this study are as follows:
\begin{enumerate}
   
    \item Which types of problems and subjects were more difficult for students?
    \item What behavioral patterns differentiate struggling students from successful students?
     \item Can artificial intelligence predict students' end-unit assignment scores with high accuracy using clickstream data?
    \item Which are the most important features for predicting students' scores?
\end{enumerate}

The rest of this paper is organized as follows: Section \ref{material_method} describes the data provided in EDMcup 2023 and discusses key aspects of our ClickTree method, including feature extraction, predictive modeling and validation. Section \ref{results} presents the results of our analysis with regards to the above research questions, including the evaluation of the ClickTree method. We discuss the implications of this study in Section \ref{implication} and provide some suggestions to improve online education in math courses. We conclude the paper and present future work in Section \ref{conclusion}.

\section{Material and Method} \label{material_method}
In this section, we will explain the ClickTree methodology used to answer RQ3. It should be noted that basic statistics (e.g. mean and variance) and visualizations (e.g. bar charts) were employed to answer the rest of the research questions.
\subsection{Data Description}
The dataset provided in the Educational Data Mining Cup 2023 \cite{edm-cup-2023} contains student clickstream data from the ASSISTments online learning platform \cite{heffernan2014assistments}. The dataset includes information about the curricula, assignments, problems, and tutoring offered to the students. ASSISTments is an online tutoring platform that provides access to math problems designed by teachers specifically for students. These problems can be solved by students during school hours or as part of their homework assignments. If students encounter difficulties in solving the problems correctly, they have the option to request hints and ask question that provide support and guidance for them to enhance their learning experience \cite{edm-cup-2023}.

In addition to assisting students in their learning journey, ASSISTments also serves as a tool to assess student performance and record their actions \cite{heffernan2014assistments,edm-cup-2023}. For this study, the dataset was collected from middle school students who used ASSISTments during the academic years of 2019-2023. The log data includes 36,296 students and 56,577 unit assignments and 57,361 problems.

For the purpose of the competition, the data consists of end-of-unit assignments given to a student, with each end-unit assignment having a list of related in-unit assignments previously completed by the same student. These in-unit assignments contain a list of problems with various tutoring options, such as requesting hints, explanations, or the full answer. Each row in the dataset corresponds to a problem within an end-unit assignment. We have a total of 226,750 rows in the training set, 225,689 in the validation set, and 124,455 in the test set. The validation set was constructed in a way that it includes half of the students from the training set and there is no overlap between the students in the training and validation set. The test set was provided by the competition. 

The data of a student in an in-unit-assignment is recorded as clickstream data, which includes several possible actions. A student can start and finish an assignment or a problem, give a correct or wrong answer, submit an open response, continue to the next problem, resume the assignment, and depending on the available tutoring options, they may request a hint, explanation, or full answer to be displayed. Additionally, students may also ask for a live chat session with a tutor or request a video explaining the skill required to solve the problem.

Furthermore, each problem is associated with its text content represented by the Bert embedding \cite{edm-cup-2023,devlin2018bert}, along with a description of its skill code as additional information. The problems can be categorized into 10 different types:
\begin{itemize}
    \item \textbf{Number:} The answer should include only a number, with no letters or mathematical symbols.
    \item \textbf{Algebraic Expressions:} The answer may consist of a combination of numbers, letters, and mathematical symbols.
    \item \textbf{Numeric Expression:} The answer should contain numbers and symbols.
    \item \textbf{Check All That Apply:} The student must select a subset of answers from a given set.
    \item \textbf{Multiple Choice:} The student must choose one answer from a set of options.
    \item \textbf{Exact Match (case sensitive, ignore case):} The student is required to submit an answer that precisely matches the correct answer. In the case of "ignore case," the distinction between lowercase and uppercase letters is not considered for the correct answer.
    \item \textbf{Exact Fraction:} The answer should be a fraction where both the numerator and denominator must match the correct answer exactly. It is important to note that even if the student's answer can be simplified to the correct fraction, it would be considered incorrect in this problem type if it doesn't exactly match the desired fraction.
    \item \textbf{Ordering:} The task is to arrange a given set in the correct order.
    \item \textbf{Ungraded Open Response:} The student is asked to upload their answer as plain text or possibly provide a voice or video explanation.
\end{itemize}

Another important feature is the ``sequence," which represents a set of problems that can be assigned to students by a teacher. Each end-unit or in-unit assignment has a sequence of problems, with each sequence having 5 levels (of which only 4 levels were used in this study). The first level specifies the curriculum, with values including the Kendall Hunt Illustrative Mathematics curriculum or the Engage New York mathematics curriculum \cite{edm-cup-2023}. The second level indicates the grade and subject, while the third level denotes the unit within the sequence. The fourth level specifies the particular subject within the unit.

\subsection{Feature Extraction}\label{fe}
The clickstream data contains raw features that reflect students' behavior, making it an excellent source for extracting information related to their performance. We have extracted various types of features from the clickstream data based on the possible action that a student can take. It should be noted that all the following features were calculated for each action (e.g. problem start, problem finish, and so on) separately.  
\begin{itemize}
    \item \textit{Assignment-level, action counts}: To determine this feature, we examined the action logs of the in-unit assignments associated with the specific end-unit assignment and calculate the count of each action (e.g. starting assignment, finishing assignment, continue and so on) performed during those in-unit assignments. The value of this feature remains the same for all problems within the same unit assignment. 

    \item \textit{Student-level, action counts}: Since a student may have completed multiple unit assignments, it is worth taking into account the count of each action performed by the student. It is important to note that the value of this feature remains constant for all rows in the dataset where the same student is involved in the assignment.
    
    \item \textit{In-unit average action counts}: For every in-unit assignment associated with an end-unit assignment, we calculated the total count of a specific action and then compute the average across all the in-unit assignments linked to that end-unit assignment.
    
    \item \textit{Problem-level average action counts}: Given that the problems within an assignment can vary in difficulty, it is important to take into account the characteristics of each problem and how students typically respond to them. This includes considering the likelihood of students providing correct or incorrect responses, as well as their tendency to request hints. Therefore, for each possible action and problem, we calculate the total count of that action for that specific problem across the entire clickstream data. Subsequently, for each in-unit assignment, we calculate the average count across all the problems within that assignment. Finally, we calculate the average count again across all the in-unit assignments associated with a particular end-unit assignment.
    
    \item \textit{Problem-weighted in-unit average}: When a student provides a correct response to a problem, the significance of this action can vary depending on the difficulty of the problem or, in other words, based on the proportion of students who typically give the correct response. To capture this significance, it is useful to incorporate a weighted measure that considers the importance of an action within an assignment.

    We defined the following function
    \[D_{\epsilon}(x)
    \begin{cases}
        \frac{1}{x}& x \neq 0 \\
        \epsilon & x =0
    \end{cases}
    \]
    Let $N_a(u,r,p)$ denote total count of action $a$ performed on problem $p$ in the in-unit-assignment $r$ within an end-unit-assignment $u$. Furthermore, let $N^{(p)}_a$ be the total count of action $a$ for the problem $p$ throughout the clickstream data. Then we have the following measure
    \begin{align*}
        D_{0}(N_a^{(p)}) N_a(u,r,p),
    \end{align*}
 The feature is calculated by averaging the measure across all problems within an in-unit assignment, and then across all in-unit assignments within a unit assignment. The measure is designed to be smaller when the action is more insignificant. This calculation takes into consideration the sufficient number of different students assigned to each problem.

    \item \textit{Problem-level performance}: For a given in-unit assignment $r$ within an end-unit assignment $u$, and a problem $p$, the following function
    \begin{align*}
        \chi(u,r,p) = D_{0}(N_\text{correct}^{(p)}) N_{\text{correct}}(u,r,p) - D_{0}(N_\text{wrong}^{(p)}) N_{\text{wrong}}(u,r,p)
    \end{align*}
    measures the performance of a student in a problem inside an assignment. To give a measure of their performance inside an in-unit assignment, sum $\chi(u,r,p)$ over all problems $p$ inside the assignment. At the end, take the average of the student's performance throughout the in-unit assignments corresponding to the end-unit assignment.
    
    \item \textit{Other features}: We have utilized a total of five categorical features in our analysis, namely, ``problem type" and ``sequence folder path levels 1 to 4". Additionally, we have included the first 32 principal components of the Bert embedding of the unit problems as additional features.
\end{itemize}

During the feature selection phase, we eliminated features that had a Pearson correlations higher than 90\%.
\subsection{Performance Predictor}
The scores of students in problems within end-of-unit assignments were predicted by the CatBoost classifier \cite{prokhorenkova2018catboost} using the novel set of features explained above. CatBoost is an efficient modification of gradient boosting methods \cite{friedman2001greedy,shyam2020competitive} that has the ability to handle categorical features without having curse of dimensionality and target leakage issues    \cite{prokhorenkova2018catboost,jabeur2021catboost,shyam2020competitive}. CatBoost consists of two main steps \cite{prokhorenkova2018catboost,shyam2020competitive,hancock2020catboost}: 1 - Pre-processing step: This step involves the efficient conversion of categorical features into numerical values known as Ordered Target Statistics (OTS) \cite{prokhorenkova2018catboost}. This conversion ensures that the categorical information is effectively incorporated into the model during training. 2 - Gradient boosting step \cite{friedman2001greedy,hancock2020catboost}: In this step, both the numerical features and the Target Statistics (TS) values of categorical features are used as input to build a gradient boosting model. This model utilizes decision trees model as the base predictors, leveraging the combined information from numerical and categorical features \cite{prokhorenkova2018catboost,hancock2020catboost}.

These steps are further elaborated in the subsequent subsections by providing a more detailed explanation of how the CatBoost classifier predicts student performance based on the set of numerical and categorical features. Comprehensive information about the CatBoost classifier and gradient boosting methodology can be found in \cite{prokhorenkova2018catboost,hancock2020catboost,shyam2020competitive}.

\subsubsection{Computing Ordered Target Statistics}
\label{subsecOTS}
In general, one-hot encoding or numerical encoding is typically employed to preprocess categorical features and use them in training models \cite{chapelle2014simple,micci2001preprocessing,seger2018investigation}. However, when dealing with high-cardinality features, such as the problem skill feature in our dataset, which has 345 different categories, one-hot encoding becomes infeasible \cite{prokhorenkova2018catboost,shyam2020competitive}. On the other hand, the TS method \cite{prokhorenkova2018catboost,hancock2020catboost} involves mapping categories to a reduced number of clusters based on their expected target value and then applying one-hot encoding to these clusters \cite{prokhorenkova2018catboost,micci2001preprocessing}. Nevertheless, this method suffers from target leakage \cite{prokhorenkova2018catboost}. To overcome this issue, CatBoost adopts a permutation-based approach to compute ordered TS that solves the issue with target leakage \cite{prokhorenkova2018catboost}. 

The computation of the OTS value for a categorical feature in the $i$th sample relies only on the targets (binary values of students' scores) of previously seen samples (samples $1, \ \cdots, \ , i-1$), and does not depend on the target of the $i$th sample \cite{prokhorenkova2018catboost}. This eliminates the problem of target leakage \cite{prokhorenkova2018catboost}. The technique that CatBoost uses to fit data to the base tree models is by making an arbitrary order \cite{prokhorenkova2018catboost,hancock2020catboost}. A permutation $\sigma$ is applied to the samples (each pair of problem and end-unit assignment) to introduce an artificial order. Let $x_k^i$ denote the $i$th categorical feature for the $k$th problem-assignment pair. The TS value for $x_{\sigma(k)}^i$ is denoted as $\hat{x}_{\sigma(k)}^i$ and is calculated using Formula \ref{ts}.

\begin{equation}\label{ts}
\hat{x}_{\sigma(k)}^i = \frac{\sum_{j=1}^{i-1}\mathds{1}_{{x_{\sigma(k)}^i = x_{\sigma(j)}^i}}.y_{\sigma(j)} + \alpha P}{\sum_{j=1}^{i-1}\mathds{1}_{{x_{\sigma(k)}^i = x_{\sigma(j)}^i}}+ \alpha }
\end{equation}

Here, $\alpha > 0$ represents a weight parameter, and $P$ denotes the average target (score) value across all training samples. To address the issue of higher variance in OTS values among preceding samples compared to later samples when using a single permutation, multiple random permutations ($s$ permutations) were applied to the problem-assignment pairs. For each permutation $\sigma_r$ (where $r \in 1,\ \cdots , \ s$), the corresponding $\hat{x}_{\sigma_r(k)}^i$ values were calculated and used in the design of the gradient boosting model \cite{prokhorenkova2018catboost,hancock2020catboost}.

Let us explain each variable in the formula with more details:

$\hat{x}_{\sigma(k)}^i$: This variable shows the output OTS value for the $i$th categorical feature (for example problem skill feature), specifically for the $k$th problem-assignment pair. It is the estimated expected target value for the sample's categorical feature based on previously seen samples.

$\sum_{j=1}^{i-1}\mathds{1}_{{x_{\sigma(k)}^i = x_{\sigma(j)}^i}}$: This part of the formula sums up the occurrences of the categorical feature $x_{\sigma(k)}^i$ being equal to the categorical feature of previous samples $x_{\sigma(j)}^i$, where $j$ ranges from 1 to $i-1$. It counts how many times the same category has been observed before the current sample.

$\mathds{1}_{x_{\sigma(k)}^i = x_{\sigma(j)}^i}$ equals to 1 if the categorical feature of the $i$th sample ($x_{\sigma(k)}^i$) is equal to the categorical feature of the $j$th previous sample ($x_{\sigma(j)}^i$), and 0 otherwise.

$y_{\sigma(j)}$: This represents the score (0 or 1) value of the $j$th pair of problem and end-unit assignment in the permutation $\sigma$. It's the actual score value corresponding to the $j$th pair of problem and end-unit assignment.

$\alpha$: A weight parameter that controls the influence of the average score (our target) value $P$ in the calculation (we set it to default value which is 0.1). It adjusts the impact of the average score value relative to the sum of individual score values.

$P$: It denotes the average score value across all pairs of problem and end-unit assignments in training data. It provides a baseline expectation for the score value, helping to normalize the contributions of individual target values in the calculation.
To explain it with an example, suppose we have the following data:

Categorical feature "Problem type" with categories: "Ordering" "Matching" and "Multi choice"
Target variable indicating student scores in pair of assignment and problem: 1 (pass) or 0 (fail).
And let us consider calculating $\hat{x}_{\sigma(1)}^1$ for the first problem-assignment pair with "Ordering" as the problem type:

$\sum_{j=1}^{i-1}\mathds{1}_{{x_{\sigma(k)}^i = x_{\sigma(j)}^i}}$: Since this is the first sample, there are no previous samples to consider, so this part is 0.
$y_{\sigma(j)}$: If, for example, the score value for the first sample is 0, then $y_{\sigma(j)} = 0$.
$\alpha$: This is a predefined parameter that can be adjusted during model training to control the influence of the average target value. The default value for this variable set to 0.1. 

\subsubsection{Building Gradient Boosting Models}
The CatBoost classifier applies a Gradient Boosting algorithm with Decision Trees (GBDT) as base predictors to predict the binary score of problem-assignment pairs \cite{dorogush2018catboost}, in our case indicating whether a student answered the problem of the end-unit assignment correctly or not. The process includes building the base tree predictors and then making the final prediction by aggregating all the base predictors.

During model construction, decision trees ($M_r$), were built based on numerical features and OTS features computed using the methodology explained in section \ref{subsecOTS}. Each tree is grown in a leaf-by-leaf manner, considering various features and combinations of features at each split that capture the relationship between categorical features such as problem skills and the target variable (student score) across different permutations, ensuring robustness in modeling. Given the infeasibility of considering all possible combinations of features, a greedy selection process was employed to select feature combinations \cite{prokhorenkova2018catboost}. In each split, the criterion for split was selected such that it minimizes the loss function defined in Formula \ref{eql2}.

\begin{equation}\label{eql2}
L2 = - \sum_{i=1}^{n}w_i \cdot (a_i - g_i)^2
\end{equation}

Here, $L2$ is the squared error loss function used in the gradient boosting model, which measures the discrepancy between the predicted score and the actual score for each problem-assignment pair. $w_i$ is the weight assigned to the $i$th problem-assignment pair, which accounts for both the output score weight ($w_i^{(y)}$) and the input score weight ($w_i^{(x)}$). It determines the importance of each problem-assignment pair in the model training process. 
For each problem-assignment pair $x_i$ with score $y_i$, the weight $w_i$ was computed by multiplying the score weight (output weight: $w_i^{(y)}$) and the problem-assignment weights (input weight: $w_i^{(x)}$). $a_i$ is the prediction of the tree for the $i$th problem-assignment pair. It represents the model's prediction of the score variable for the given problem-assignment input. $g_i$ is the gradient of the loss function based on the tree prediction. It shows the sensitivity of the loss function to changes in the predicted scores.
The problem-assignment weights were randomly assigned using Bayesian bootstrap \cite{rubin1981bayesian} with a bagging temperature of 0.2, and the output class weights were calculated based on Formula \ref{classweight}.

\begin{equation}\label{classweight}
w_i^{(y)} = \sqrt{\frac{\max_{c=1}^K(\sum_{y_j = c}w_j^{(x)})}{\sum_{y_j = y_i}w_j^{(x)}}}
\end{equation}

where $w_i^{(y)}$ is score weight assigned to the $i$th problem-assignment pair, which accounts for the distribution of problem-assignment pairs across different classes. It aims to balance the contribution of two score classes (0 and 1) in the model training process. $K$ is the number of classes in the score variable that is equal to 2. $w_j^{(x)}$ is the input weight assigned to the $j$th problem-assignment pair. As mentioned above, it reflects the importance of each assignment-problem pair in the model training process. $y_i$ is target score for the $i$th problem-assignment pair, indicating whether the pair was solved correctly or not by the student.

Each tree is constructed leaf by leaf until it reaches the maximum depth or the maximum number of leaves, whichever occurs sooner. The overall loss function for the model is cross-entropy (CE) \cite{de2005tutorial}, as defined in Formula \ref{eqloss}.

\begin{equation}\label{eqloss}
CE = -\dfrac{\sum_{i=1}^{n}w_i (y_i \log (p_i) + (1 - y_i) \log (1-p_i))}{ \sum_{i=1}^{n}w_i }
\end{equation}

Here, $y_i$ represents whether the $i$th problem-assignment pair is solved correctly or not, and $p_i$ is the prediction of the model. The model was trained using stochastic gradient Langevin boosting with posterior sampling \cite{ustimenko2021sglb}.

After constructing decision trees $M_1$, ..., $M_s$ based on OTS values computed using permutations $\sigma_1$, ..., $\sigma_s$, an additional permutation $\sigma_{*}$ was used to compute OTS values for the final prediction and to determine the leaves in the trees. The final prediction for each problem-assignment pair is computed by boosting the predictions of all the models. To avoid overfitting, early stopping is enabled, meaning that the model stops learning when its performance on the validation data does not improve. This approach allows saving the best-performing model on the validation data, which is subsequently used as the final model for predicting problem-assignment results in the test data.

\subsection{Validation}
The validation set was constructed to ensure that it includes half of the students from the dataset, and there is no overlap between the students in the training and validation sets. Consequently, every row in the validation set corresponds to a student who is not present in the training data.

To optimize the performance of the CatBoost classifier, we performed hyperparameter tuning within specified ranges. The hyperparameters we focused on were as follows: `depth' with options [3, 1, 2, 6, 4, 5, 7, 8, 9, 10], `iterations' with options [250, 100, 500, 1000], `learning\_rate' with options [0.03, 0.001, 0.01, 0.1, 0.2, 0.3], `l2\_leaf\_reg' with options [3, 1, 5, 10, 100], 
`bagging\_temperature' with options [0.03, 0.09, 0.25, 0.75], `random\_strength' with options [0.2, 0.5, 0.8].
The objective of this tuning process was to increase the AUC value on the validation data. After the initial tuning, we manually fine-tuned the selected hyperparameters.

The best hyperparameters that were ultimately chosen for the model are as follows: n\_iterations = 5000, learning\_rate = 0.01, use\_best\_model = True, l2\_leaf\_reg = 200, depth = 10, score\_function = 'L2', langevin = True, grow\_policy = 'Lossguide', auto\_class\_weights = 'SqrtBalanced', eval\_metric = 'AUC', posterior\_sampling = True, bootstrap\_type = 'Bayesian', bagging\_temperature = 0.2, sampling\_unit = 'Object', early\_stopping\_rounds = 100. Other parameters are set as deafult.

The implementation was carried out using Python 3, along with the following libraries: numpy 1.19.5, pandas 1.3.4, catboost 1.1.1, xgboost 1.7.5, and scikit-learn 0.24.2. The code was executed on a computer with 8 CPU cores and 16 GB of RAM. The code and implementations are available at: \newline \url{https://www.kaggle.com/code/nargesrohani/clicktree/notebook}

\section{Results} \label{results}
In this section, we present the results of our analysis regarding the research questions, as well as the evaluation of the ClickTree method. We also compare the performance of the method with other traditional machine learning algorithms.

\subsection{Which problems were more difficult for students?}
\begin{figure}
    \centering
    \includegraphics[scale=0.63]{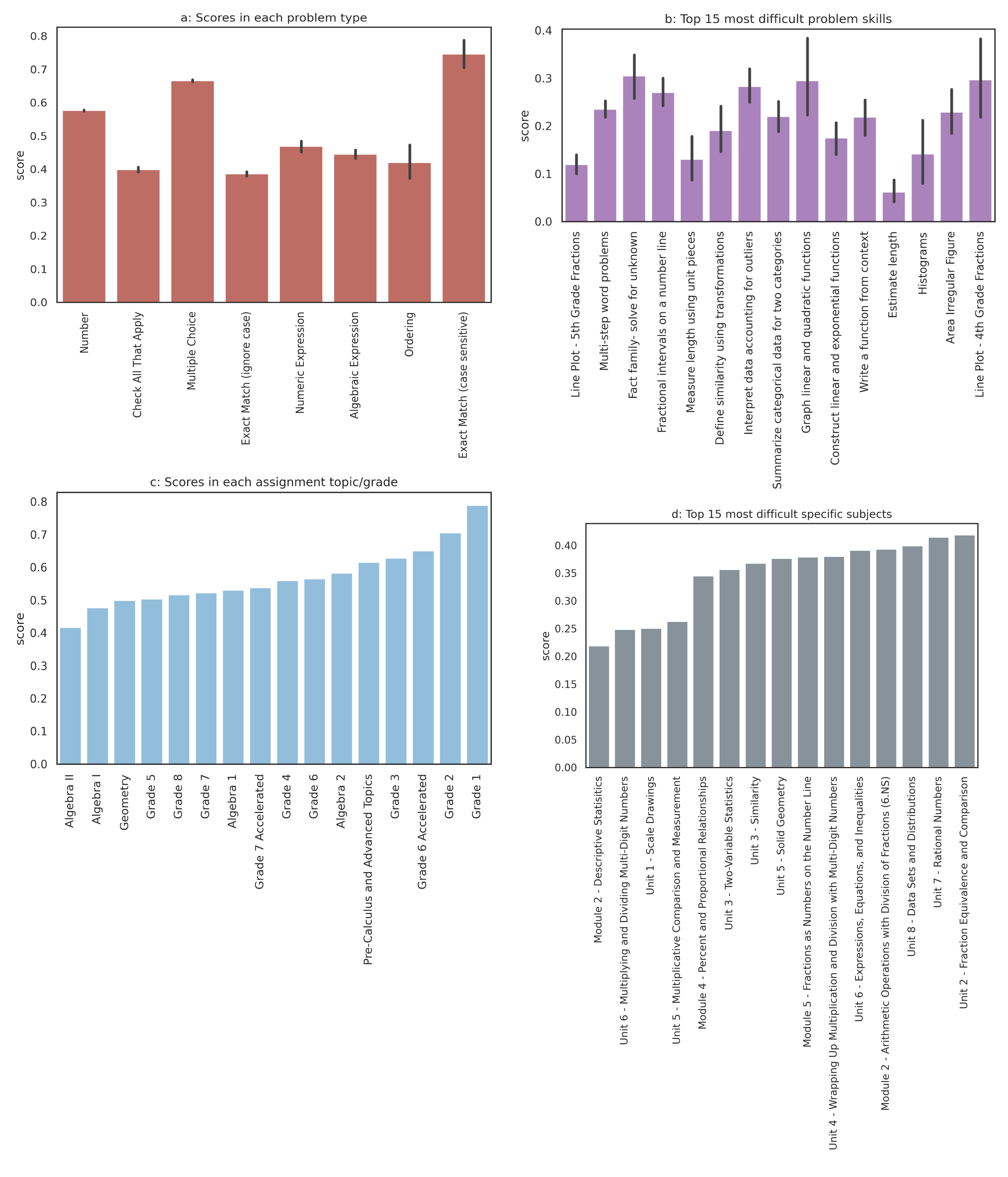}
    \caption{a: Average scores in different types of problems. b: 15 most difficult problem skills with more than 100 occurrences in the whole dataset (problem skill indicates the skill required for solving a problem ). c: Average scores among different topic/grade of assignments. d: 15 most difficult problem subjects with more than 100 occurrences in the whole dataset}
    \label{fig_1}
\end{figure}

Among the different types of problems, `Exact match (ignore case)' with an average score of 0.38, `Check all that apply' with an average score of 0.40, and `Ordering' with an average score of 0.42 had lower average scores compared to other types of problems (Figure \ref{fig_1}.a). This indicates that they were more difficult for students.

Figure \ref{fig_1}.b displays the 15 most challenging problems based on the skills required. Problems that require estimating length (average score of 0.06), Line plot - 5th grade fraction (average score of 0.12), measuring length using unit pieces (average score of 0.13), and histograms (average score of 0.14) were found to be more difficult compared to other problem skills.

Based on Figure \ref{fig_1}.c, which presents the average scores among different sequence levels/subjects of the assignments, Algebra II with an average score of 0.42, Algebra I with an average score of 0.48, and Geometry with an average score of 0.50, were more challenging for students.

Furthermore, when examining subtopic of the problems in Figure \ref{fig_1}.d, Module 2 - descriptive statistics (average score of 0.22), Unit 6 - multiplying and dividing multi-digit numbers (average score of 0.25), Unit 1 - scale drawings (average score of 0.25), and Unit 5 - multiplicative comparison and measurement (average score of 0.26) were found to be more difficult for students.

\subsection{Learning behavior of successful students vs struggling students}
The results show that students who answered the end-unit assignment problems correctly (i.e. successful students), had a significantly higher number of `continue selected', `correct response', `open response', `problem finished', and `problem started' actions in in-unit assignments compared to the ones struggling when answering the in-unit assignments (Table \ref{tableee}). Students who could not answer the end-unit assignment problems correctly (i.e. struggling students) had a lower number of actions in the above categories, and they also had a higher number of requests for hints, answers, and explanations. This implies that students who struggle can be identified during their interactions with the in-unit assignments before they take the final unit assignment. Hence, if teachers provide them with suitable support (e.g. explanations and tutoring) they could potentially answer the final assignments more successfully. 
\begin{table}[h]
  \caption{The average in-unit action counts for each group of students. A successful student is one who answered the problem correctly, while a struggling student is the one who could not answer the problem correctly. Actions with a significant statistical difference between struggling and successful groups are highlighted in bold.}\vspace*{1ex}
  \label{tab:1}
  \centering
  \begin{tabular}{| l | l | l |}
    \hline
    \multicolumn{1}{|c|}{\textbf{Action}} & \multicolumn{1}{c|}{\textbf{Struggling students}}& \multicolumn{1}{c|}{\textbf{Successful students}} \\
    \hline
        \textbf{answer requested} &18.16&11.28\\
    assignment finished &11.10&12.32\\
    assignment resumed &7.97&
8.09
\\
    assignment started&13.56&
14.38
\\
    \textbf{continue selected}&101.18&
117.97
\\
    \textbf{correct response}&78.43&
93.16
\\
    \textbf{explanation requested}&0.61&
0.33
\\
    \textbf{hint requested}&2.24&
1.31
\\
    \textbf{open response}&33.74&
37.14\\
    \textbf{problem finished}&112.48&
130.51\\
    \textbf{problem started}&114.88&132.51\\
    \textbf{wrong response}&40.61&33.90\\

    \hline
  \end{tabular}
  \label{tableee}
\end{table}

\subsection{ClickTree method evaluation}
\begin{figure}
    \centering
    \includegraphics[scale=0.7]{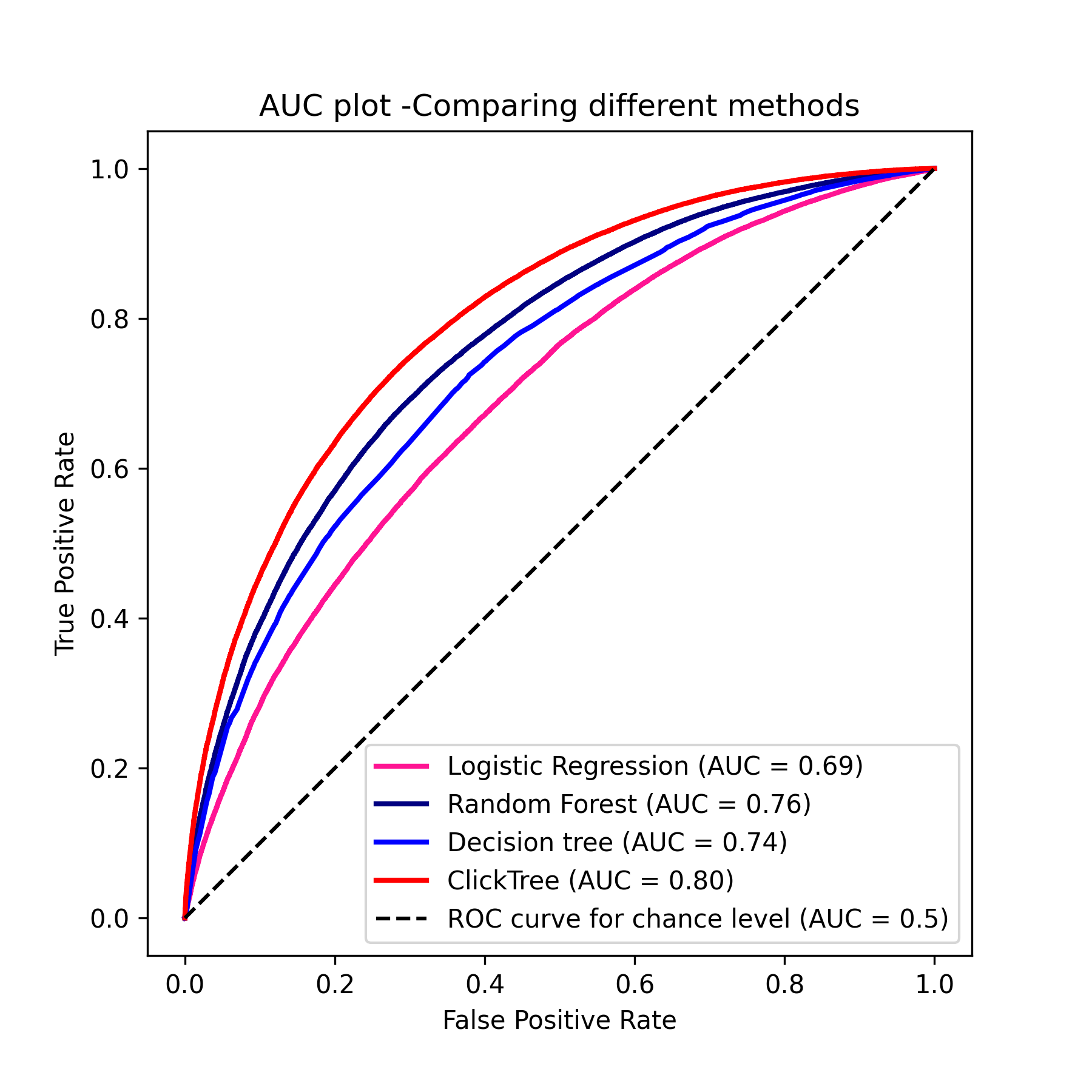}
    \caption{The AUC values of the different methods on the validation data}
    \label{fig:auc}
\end{figure}

The ClickTree method achieved an AUC of 0.78844 on the test data and ranked second in the EDM Cup 2023. Before using the CatBoost classifier, we applied different classification methods to the extracted features in order to find a classifier with higher accuracy in predicting students' end-unit scores. Figure \ref{fig:auc} shows the AUC values of the proposed method, which uses the CatbBoost classifier as a predictor, compared to other traditional machine learning methods. The ClickTree method achieved an AUC of 80\% on the validation data, while Random Forest, Decision Tree, and Logistic Regression achieved AUC values of 76\%, 74\%, and 69\%, respectively. One can conclude that tree-based methods were more appropriate for the dataset. It should be mentioned that since we had categorical features, the CatBoost classifier, which can handle categorical data, seems to be more suitable and accurate. To use categorical features in the classification by Random Forest and Decision Tree, we employed the Label Encoding approach to convert the categorical values into numerical values.

\subsection{Feature Importance}
\begin{figure}
    \centering
    \includegraphics[scale=0.7]{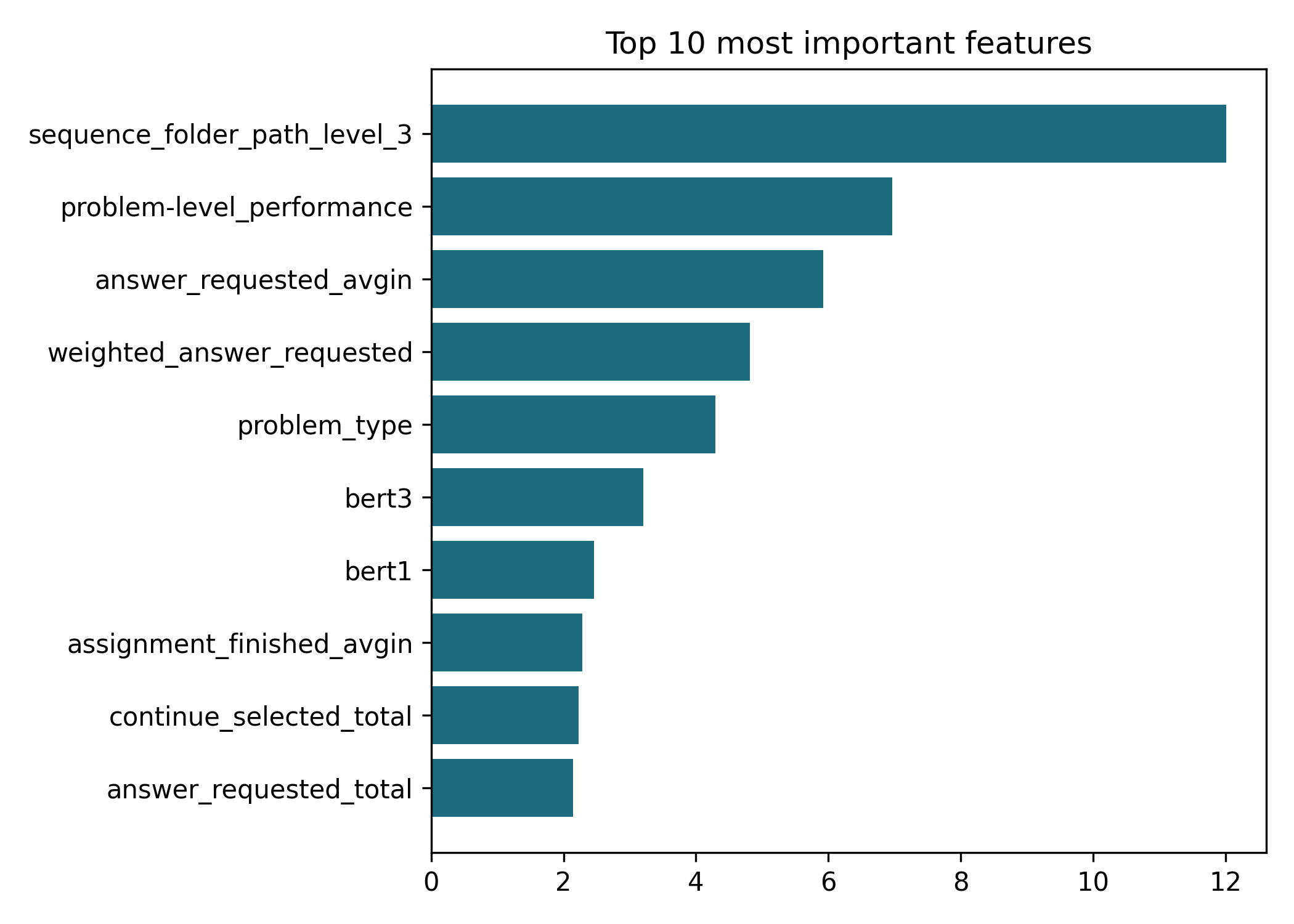}
    \caption{Top 10 most important features for the CatBoost classifier based on the information gain.}
    \label{fig:important}
\end{figure}

Upon analysing the importance of each feature in increasing the ClickTree predictive model accuracy, we identified the 10 most important ones (see Figure \ref{fig:important}). The most important feature is sequence\_folder\_path\_level\_3, which represents the subject of the assignment problem. The next most important features are problem level performance, average answer request, weighted answer request, problem type, BERT features calculated from the questions' text, average assignment finished, total continue of the assignments, and total answer request, respectively (features explained in Section \ref{fe}).

\section{Implications and Recommendations}\label{implication}
The results revealed that certain types of problems were more challenging for students. Specifically, the problem types ``Exact match (ignore case)", ``Check everything relevant", and ``Ordering" had lower average scores, indicating greater difficulty. Targeted support can be provided to address these challenges. For example, for the type of ``Exact matching (ignore case)" problems, additional practice or guided examples can be provided to improve student understanding of required skills for this problem type.

Also, problems that required skills such as estimating lengths, line plots, measuring lengths by unit pieces, and histograms were found more difficult by students. Also, some topics were identified as more difficult for students (such as Algebra II, Algebra I, and Geometry). Therefore, new teaching strategies and resources can be designed to teach these topics more effectively. For example, offering creative examples, practice, and interactive activities specifically focusing on the difficult topics can help facilitate student training. Another suggestion is to provide clear explanations and step-by-step instructions for the topics. This insight may enable educators to devote more attention and instructional resources to the challenging subjects and sub-subjects, ensuring that students receive appropriate support and instruction where needed most.

Students’ behavior analysis revealed notable differences between successful and struggling students. Successful students, who answered the end-unit assignment problems correctly, exhibited high engagement levels. On the other hand, struggling students requested more hint, answers, and explanations while exhibiting lower overall engagement. This finding suggests that struggling students can be identified early on based on their interactions with in-unit assignments. By providing timely personalized support and interventions for these students can improve their performance in end-unit assignments. Additionally, offering self-directed learning strategies and helpful online tutorials or educational apps to support students' independent learning could be beneficial.

To summarize, the insights derived by this study provide valuable knowledge and guidance for educators and course instructors. They can apply this knowledge to math courses to enhance instructional strategies, prioritize specific problem areas, tailor support and interventions to struggling students, and employ machine learning techniques such as ClickTree for accurate performance prediction.

\section{Conclusion and Future Work} \label{conclusion}
We have developed ClickTree, a method for predicting students' performance in math assignments based on their clickstream data. ClickTree calculates a novel set of features at different levels, including the problem-level, student-level, and assignment level. After that, it utilizes the CatBoost classifier to predict students' scores in answering the final unit assignment problems. The developed method achieved an accuracy of approximately 79\% and ranked second in EDMcup23. During the training of the method, having a validation set that demonstrates test data was crucial. The validation set was carefully selected to include new students, and our results show that tree-based methods with a boosting approach had higher accuracy in the dataset. The inclusion of multiple levels of features was effective in improving model accuracy, and the Bert vectors, which represent the text of the questions, were also important features for prediction.

Our analysis also reveals that students who started and completed more problems, as well as those who answered more questions correctly during the in-unit assignment, were more likely to answer the final assignment of the unit correctly. However, struggling students had a higher rate of requesting answers, explanations, or hints for the problems.

We also investigated the types and subjects of problems that were more challenging for students. The results demonstrate that the type of problems where a student must select a subset of answers from a given set (i.e. `Check all that apply') was more difficult for students, and topics such as Algebra II, descriptive statistics, and multiplying and dividing multi-digit numbers posed more difficulty for students.

By considering the insights provided in this paper, course instructors could focus on improving the course and assignments in the subjects and problem areas that were difficult for students. Additionally, they could provide students with more tutoring in subjects that could help them learn better. Moreover, the identification of struggling students using ClickTree could be utilized to provide timely and personalized help to students. Related work on utilizing learning analytics to implement effective student interventions could be useful here \cite{smith2012predictive,wong2017learning}. 

However, it is important to note that the generalizability of the developed method should be checked using other courses and datasets in future research. Also, the calculation of features that can reflect students' learning processes might improve the method accuracy. Process mining methods that can extract the sequence of actions and transitions between different in-unit assignments from click-stream data are potential approach to explore in future work. 

\section{Acknowledgments}
We would like to thank the Educational Data Mining conference for organising EDM Cup 2023. This work was supported by the Medical Research Council [grant number MR/N013166/1].
\section{Author Contributions}
NR and BR contributed to the development of the method. NR and BR generated the results and implemented the method, while AM helped in interpretation of the results. NR and BR wrote the first draft of the paper and AM assisted in enhancing the written work. The authors all read the final version of the paper and approved it.
\nocite{*}

\bibliographystyle{acmtrans}
\bibliography{ref}

\end{document}